\documentclass{pazh}
\usepackage{graphicx}
\usepackage{latexsym}

\begin{document}

\title{The Gas Content in Galactic Disks: Correlation with Kinematics}

\author{A.V. Zasov$^1$ \and A.A. Smirnova$^2$}

\institute{ Sternberg Astronomical Institute, Universitetskii pr. 13, Moscow, 119992 Russia
\and Special Astrophysical Observatory of RAS, Nizhnii Arkhyz, Karachai-Cherkessian
Republic, 357147 Russia}

\date{Received -- June~22, 2004}

\offprints{Anatoly Zasov  \email{zasov@sai.msu.ru}}

\titlerunning{The Gas Content in Galactic Disks}
\authorrunning{ Zasov \& Smirnova}

\abstract{ We consider the relationship between the total HI mass
in late-type galaxies and the kinematic properties of their
disks. The mass $M_{HI}$ for galaxies with a wide variety of
properties, from dwarf dIrr galaxies with active star formation
to giant low-brightness galaxies, is shown to correlate with the
product $V_{c}R_{0}$ ($V_{c}$ is the rotational velocity, and
$R_{0}$ is the radial photometric disks scale length), which
characterizes the specific angular momentum of the disk. This
relationship, along with the anticorrelation between the relative
mass of $HI$ in a galaxy and  $V_{c}$, can be explained in terms
of the previously made assumption that the gas density in the
disks of most galaxies is maintained at a level close to the
threshold (marginal) stability of a gaseous layer to local
gravitational perturbations. In this case, the regulation
mechanism of the star formation rate associated with the growth
of local gravitational instability in the gaseous layer must play
a crucial role in the evolution of the gas content in the
galactic disk.} \maketitle

\section{INTRODUCTION}

Elucidating the mechanisms that determine the
current gas content in galactic disks is an outstanding
problem. Several processes that are capable to change significantly
the total mass of the interstellar medium in galaxies over their lifetimes are known.
These include star formation, galactic wind, the ejection
of matter by evolved stars, the accretion of intergalactic
gas, and the absorption of gas-containing
companion galaxies. Undoubtedly, the efficiency of
these processes and the extent to which they are
balanced vary greatly from one galaxy to another, so the
total mass of the gas in a galaxy can both decrease
and increase with time during certain evolutionary
periods. Still, the main mechanism that determines
the current mass of the gas is certainly its consumption for
star formation: if we proceed from the current star
formation rates (SFRs), then, in most cases, the gas
depletion time scale $t_{c}$ proves to be much smaller
than the Hubble age of galaxies (see, e.g., Devereux
and Hameed 1997; Bendo et al. 2002; Zasov and
Bizyaev 1994; Zasov 1995).

As was first shown by Larson and Tinsley (1978),
the color differences between galaxies can be well
explained in terms of a difference in the SFR decay
rates at the same age of the systems: in redder
galaxies, star formation was more intense and has
been almost completed. From this point of view, the
relative amount of the gas left in the galactic disk
must be determined by the star formation efficiency in
them (the current SFR per unit gas mass), which is
higher in systems with more favorable star formation
conditions. Undoubtedly, the current SFR must be
affected by the rotational velocity of the gas via both
the angular velocity of spiral density waves in the
disk and the formation conditions of large-scale gas
condensations, which are definitely different in fast
and slowly rotating disks.

Actually, however, the situation proves to be not
so simple. The current star formation efficiency (or its
reciprocal, the gas depletion time scale) was found
to correlate rather weakly with other parameters of
galaxies. It weakly correlates with the gas mass-to-luminosity
ratio for galaxies (Zasov 1995) and shows
no clear correlation with the morphological type, the
luminosity (Boselli et al. 2002), and the rotational velocity
or color of spiral galaxies (Boissier et al. 2001).
Both the observed star formation efficiency and the
relative gas content in the disk, which is defined as
the ratio $M_{HI}/L$, can differ in galaxies of the same
morphological type or in galaxies with the same color
index by more than an order of magnitude (Verheijen
and Sancisi 2001; Boissier et al. 2001).

 Hence, it would be natural to expect the total mass of
the gas in the disk, which is a derivative of many
factors and primarily of the SFR history, to be also
insensitive to other parameters of disk galaxies. Nevertheless,
several simple relationships prove to exist
between the gas mass and galactic disk parameters.

(1) \textit{The relationship between $M_{HI}$ and the
galaxy size}. Although the mass fraction of the
gas in a galaxy ($M_{gas}/M_{tot}$) does not correlate
with the linear size of the galaxy (McGaugh and
de Blok 1997), observations reveal a close correlation
between the total gas mass $M_{HI}$ and the diameter D
exhibited by all types of disk galaxies, except the
earliest types (S0--Sab) (Hewitt et al. 1983; Broeils
and Rhee 1997; Becker et al. 1988; Martin 1998).
The form of the relationship is $M_{HI} \sim D^{n}$, where
$n \approx 1.8-2$, implying that the mean neutral hydrogen
surface density $\langle \sigma_{HI}\rangle$ is approximately constant in
various galaxies. If the early-type galaxies are excluded,
then $\langle \sigma_{HI}\rangle$ undergoes virtually no systematic
changes with galaxy morphological type T and rotational
velocity (Karachentsev et al. 1999a, 2004).
Low- and high-surface-brightness galaxies also lie
on the same $M_{HI}(D)$ relationship (Verheijen and
Sancisi 2001).

To explain the approximate constancy of $\langle \sigma_{HI}
\rangle$, Shaya and Federman (1987) suggested the transition of
the gas to a molecular state when the ionizing radiation is
screened by an HI layer where its surface density exceeds a
certain threshold value. This mechanism can indeed partly or
completely explain the slower decrease in the azimuthally
averaged surface density $\sigma_{HI}$(R) compared to the
molecular gas surface density, but the constancy of the mean gas
surface density in galaxies by no means follows from it.
Moreover, if the total atomic and molecular gas surface density
$\langle \sigma_{HI+H_{2}}\rangle$ is taken instead of $\langle
\sigma_{HI}\rangle$, then it will prove to also change little
along the sequence of morphological types, which by no means
follows from the suggested mechanism. This is clearly illustrated
by Fig.1, in which the total gas mass (including the $HI$,
$H_{2}$, and He masses) is plotted against the optical galaxy
diameter for late-type (T $>$ 3) galaxies using data from the
catalog by Bettoni et al. (2003) (below designated as CISM).
Including earlier-type galaxies increases the scatter of points
without changing significantly the slope of the relationship.

The correlation of the HI mass with the disk size is particularly
pronounced when the radial disks scale length $R_{0}$ is
substituted for the isophotal diameter (Swaters et al. 2002) or
when $D_{HI} (= 2R_{HI})$, the size of the gaseous disk bounded
by a fairly low threshold gas surface density ($\sigma_{HI} =
1M_{\odot} pc^{-2}$) is used as the diameter, within which almost
of the HI mass is contained (Verheijen and Sancisi 2001) .

\begin{figure}
\centering
\includegraphics[width=7cm, angle=-90 ]{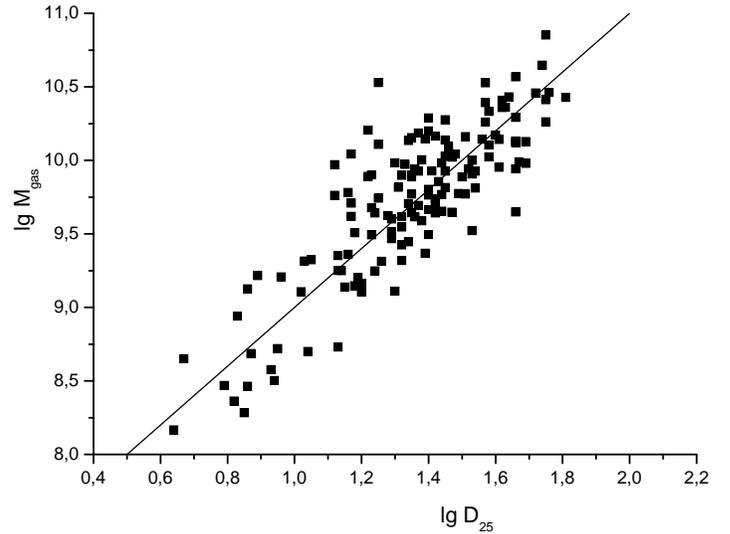}
\caption{Total mass of the interstellar gas (in solar units)
versus optical diameter $D_{25}$ (in kpc) for late-type (T $>$ 3)
galaxies (CISM data).} \label{array}
\end{figure}

(2) \textit{The decrease in the relative gas content
$M_{HI}/L$ with increasing surface brightness of late type
galaxies} (Karachentsev et al. 1999a; Mc-
Gaugh and de Blok1997; Swaters et al. 2002). This
decrease can be easily shown to be a direct result of
the relationship $M_{HI} \sim R^{2}_{0}$. Indeed, since the disk
luminosity L is proportional to $I_{0}R^{2}_{0}$, where $I_{0}$ is the
disk surface brightness extrapolated to the center, the
relationship $M_{HI} \sim R^{2}_{0}$ leads to inverse proportionality
between $M_{HI}/L$ and $I_{0}$. For galaxies, whose
brightness  differ by five magnitudes, the $M_{HI}/L$
ratios must differ by two orders of magnitude, in
excellent agreement with the observations of nearby
late-type galaxies (see Fig. 5 from Karachentsev et
al. 1999a).

(3) \textit{The relationship between the galaxy rotational
velocity $V_{c}$ and the relative gas mass $M_{HI}/M_{25}$.}
Here, $M_{25}$ is the indicative mass of the galaxy within its
optical radius equal to $V^{2}_{c}D/2G$, where $V_{c}$ can be
determined from the HI line width corrected for the disk
inclination. The faster the galaxy rotation, the lower, on
average, its relative gas content (Karachentsev et al. 1999a,
2004; Boissier et al. 2001). As an illustration, Fig. 2 shows the
$log(M_{25}/M_{HI})$ - $log(V_{c})$ diagram based on the data
from the catalog of nearby galaxies by Karachentsev et al. (2004)
(where the observational selection effects are probably at a
minimum) for Sbc and later-type galaxies. Such a relationship
cannot be a simple reflection of different compression ratios of
the gas when it passes through spiral density waves, since
irregular (including dwarf) galaxies, which constitute a majority
among the galaxies of the catalog, also obey it.

\textit{(4) The relationship between $M_{HI}$ and the disk
specific angular momentum $V_{c}D$ }(Zasov 1974; Zasov and
Rubtsova 1989). The higher the specific angular momentum, the
larger the amount of gas in the galaxy. As we show below, this
relationship, which has a simple physical interpretation, is
probably the key one. In this paper, we restrict our analysis to
the total mass of the gas in late-type galaxies. We show that the
relationships listed above can be explained by the fact that the
bulk of the gas in galaxies have a surface density close to its
threshold value $\sigma_{c}$ for a gravitationally stable gaseous
layer, and that the galaxy diameter correlates with the disk
rotational velocity.

\begin{figure}
\centering
\includegraphics[width=7cm, angle=-90 ]{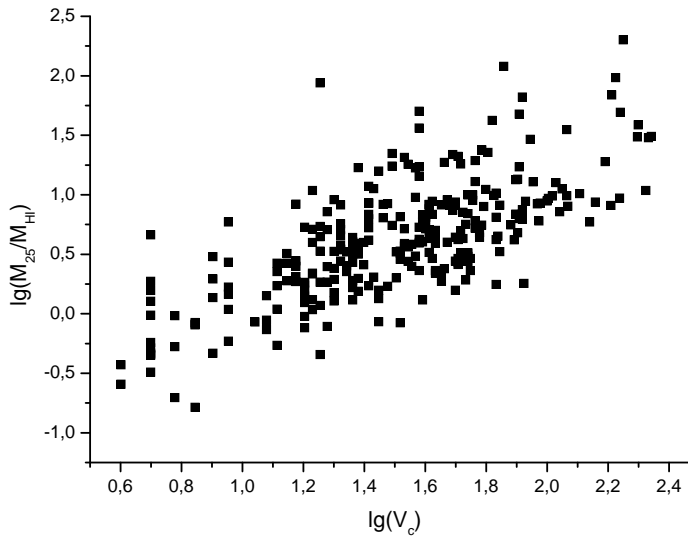}
\caption{Ratio of the total (indicative) galaxy mass to the total
HI mass (in solar units) versus rotational velocity, as
constructed for late-type (T $>$ 3) galaxies using data from the
catalog of nearby galaxies by Karachentsev et al. (2004).}
\label{array}
\end{figure}

\section{THE GAS MASS - SPECIFIC ANGULAR
MOMENTUM RELATIONSHIP FOR A
MARGINALLY STABLE GASEOUS LAYER}
Beginning with the classic paper by Quirk(1972),
various authors have considered the relationship between
the radial gas density distribution in late-type
galaxies and the distribution of the critical density $\langle \sigma_{c} \rangle$
for the growth of local gravitational perturbations. In
the simplest case of a thin gaseous disk in the gravitational
field of a galaxy, it is defined by the equation

\begin{equation}
\sigma_{c}= c_{g}\kappa/Q_{T}\pi G
\end{equation}

Here $c_{g}$ is the one-dimensional velocity dispersion of
gaseous clouds (6-8 km s$^{1}$), which is commonly assumed to be
constant and independent of the morphological type and the
galactocentric distance, as direct HI and CO measurements show
(except for the galactic circumnuclear regions) (see, e.g., Lewis
1984; Combes et al. 1997); $\kappa\ $ is the epicyclic frequency;
and $Q_{T}>1$ is the dimensionless (Toomre) parameter of stability
to arbitrarily perturbations in the plane of the disk. In
general, its value depends on the shape of the rotation curve and
the mass distribution in the disk can be determined analytically
or by numerical simulations of the growth of disk instability.
For purely radial perturbations of a thin disk, $Q_{T}$=1. The
existing theoretical $Q_{T}$ estimates that include nonradial
perturbations were obtained only for simplified models of gaseous
disks, and their values lie within the range $Q_{T}$=1.2-1.7 (see
Morozov 1985; Polyachenko et al. 1997; Kim and Ostriker 2001; and
references therein). The latter authors took into account the
magnetic field of the interstellar medium, which does not change
the results significantly \footnote{The critical density (1) was
determined under the assumption of a one-component gaseous disk.
The presence of a stellar disk makes the gaseous disk slightly
less stable. However, its effect is quantitatively small, since
the velocity dispersion of the old disk stars is much higher than
that of the gas and since the density of the stellar disk is very
low in the outer regions where their values could be comparable.
Boissier et al. (2003) showed for several spiral galaxies that
including the stellar disk reduces the local critical mass by
10-15\% and, as it is natural to expect, the effect of the
stellar disk decreases with increasing galactocentric distance.}.

If the gravitational stability of the outer gaseous disk
determined for the azimuthally averaged gas density (where
$\sigma < \sigma_{c}$) is assumed to be responsible for the sharp
reduction in SFR at a certain galactocentric distance $R_{c}$
where $\langle\sigma_{HI}\rangle = \sigma_{c}$, then a comparison
of $R_{c}$ with the radius of the outer boundary of the
distribution of HII regions corresponds most closely to
$c_{g}/Q_{T} \approx$4 km s$^{-1}$ (Martin and Kennicutt 2001);
whence it follows that $Q_{T} \approx 1.5-2$ at $c_{g} \approx 6-8
km s^{-1}$.

 Although the question of how closely the boundary
of active star formation in galactic disks corresponds to the
radius $R_{c}$ at which the gas density becomes equal to
$\sigma_{c}$ is debatable, and this condition is definitely not
satisfied in all of the galaxies studied, the it is evident, that
the azimutally averaged density $\sigma_{HI}(R)$ or
$\sigma_{HI+H_{2}}(R)$ in most of the S and Irr galaxies is close
to $\sigma_{c}(R)$ over a wide range of galactocentric distances,
differing from it by no more than a factor of 2 \footnote{The gas
density is most often higher in the inner galactic region and
lower in the outer disk regions than its critical value.} (Zasov
and Simakov 1989; Martin and Kennicutt 2001; Wong and Blitz 2002;
Boissier et al. 2001; Hunter et al. 1998). Including the
molecular gas poses a serious problem in estimating the gas mass
in the disk. This is because, first, the dependence of the
CO-H$_{2}$ conversion factor on chemical abundances is known
poorly, and, second, direct CO measurements are generally
restricted only to the inner galactic region, so when the total
mass of the molecular gas is estimated, one has to extrapolate
its density to large galactocentric distances. However, the
latter is unlikely to play a significant role, since the H$_{2}$
fraction decreases with R (Wong and Blitz 2002).

The situation is alleviated by the fact that the bulk of the
observed gas is in the form of HI in most of the galaxies (at
least in the late-type ones): the molecular hydrogen mass
M$_{H_{2}}$, on average, accounts for about 15\% of the mass
M$_{HI}$ (Casoli et al. 1998; Boselli et al. 2002). Note,
however, that this value may prove to be slightly underestimated:
according to the CISM, which combines the H$_{2}$ estimates
obtained by various authors by assuming the conversion factor to
be constant (2.3x10$^{20}$ mol K$^{-1}$ km s$^{-1}$), the
molecular gas (taking into account helium) accounts for a
slightly higher mass fraction ($\sim$40\% of the HI mass), except
for the Sd-Irr galaxies that contain a very small amount of
molecular gas (Bettoni et al. 2003).

In any case, the inclusion of the molecular gas increases
the total mass of the cold gas in spiral galaxies, on
average, by no more than a factor of 1.5 (although this
factor could be much larger in some of the galaxies),
which justifies using the HI mass estimate to characterize
the total amount of the gas. The predominance
of the atomic gas makes it easier to analyze
the evolution of the gaseous component of the disk,
since the mass of the molecular gas is known for
a much smaller number of galaxies than the HI
mass. Here, we do not consider the possibility that
very cold and, hence, unobservable H$_{2}$  cloudlets in
galactic disks can form a layer with such a high total
surface density that they produce the effect of hidden
mass by contributing significantly to the disk mass
(Combes and Pfenniger 1997). This hypothesis encounters
serious difficulties in analyzing the gravitational
stability of a gaseous layer (Elmegreen 1995) or
in dynamical mass estimations for the disks of spiral
galaxies (Kranz et al. 2003; Zasov et al. 2004), which
do no leave much space for dark matter in the disks.

Let us estimate the gas mass M$_{c}$ in a disk, expected if the gas
density is close to the critical value over its entire
length.

  Let the gas density everywhere from the center
(R=0) to the radius R$_{HI}$ within which almost all of the HI
mass is contained be defined by Eq. (1). We will approximate the
circular velocity by a simple function, $V (R) =
V_{c}(R/R_{HI})^{n}$, where $V_{c}$ is the rotational velocity on
the periphery of the galaxy ($R \sim R_{HI}$), and the constant
$n$ can have values for different galaxies from about zero (a
plateau on the rotation curve) to unity (rigid-body  rotation).
In the inner regions of spiral galaxies, the rotation curve is
generally more complex in shape, but the bulk of the HI is
located in regions far from the center, where $V_{c}(R) \sim
const$ $(n \sim 0)$. In irregular galaxies, $0 < n < 1$, while for
the least massive systems, the parameter $n$ could be close to
unity. For the epicyclic frequency, we have

\begin{equation}
\kappa = 2\Omega[1+(n-1)/2]^{1/2}
\end{equation}

where $\Omega$ is the angular velocity of circular rotation of
the disk; whence it follows that the critical gas mass
is

\begin{equation}
M^{c}_{gas} =\int\limits_0^{R_{HI}} {2\pi R\sigma_{c}(R)dR}
\end{equation}
\begin{equation}
\frac{2^{2/3}}{G}\frac{c_{g}}{Q_{T}}(1+n)^{1/2}V_{c}R_{HI}
\end{equation}

The total mass of the gas is related to the masses of
the atomic and molecular gases by

\begin{equation}
M_{gas} = 1.4(M_{HI} +M_{H2}) = \eta^{-1}M_{HI},
\end{equation}

where $\eta^{-1}\sim1.4-2$ is a coefficient that
characterizes the fraction of the molecular gas, helium, and
heavier elements in the total mass of the gas (the first value
corresponds to a negligible mass of the molecular gas). It
follows from (3) and (4) that

\begin{equation}
M_{HI}^{c}=\eta M_{gas}^{c}=\eta2^{3/2}K(1+n)^{1/2}V_{c}R_{HI}/G
\end{equation}

where $K \equiv c_{g}/Q_{T}$. Given the uncertainties in the
$Q_{T}$ and $c_{g}$ estimates (see above), $K$ may be assumed to
lie within the range 3.5 - 6.5 km s$^{-1}$.

 Since the resulting estimate depends weakly on
the parameter $n$ (the largest uncertainty is associated
with the ratio $c_{g}/Q_{T}$ and with the assumption of its
constancy along the radius), we assume below that
$n = 0$, which corresponds to the same rotational velocity
at all $R$. In this case,

\begin{equation}
M^{c}_{HI} = 2^{3/2}\eta\frac{K}{G}V_{c}R_{HI}.
\end{equation}

A relation similar to (5) (but with a different numerical
value of the proportionality coefficient) can
also be written for the case where the photometric
radius $R_{25}$, which, on average, is a factor of 1.7-1.8
smaller than $R_{HI}$ for both spiral and irregular galaxies
(Broeils and Rhee 1997; Swaters et al. 2002), is
used instead of $R_{HI}$ as the upper integration limit
in Eq. (3). Therefore, Eq. (6), to within numerical
coefficients, describes the previously reached conclusion
(Zasov 1974; Zasov and Rubtsova 1989) that
the total mass of the gas at its threshold density is
proportional to its specific angular momentum $DV_{c}$,
where $D$ is the optical diameter of the galaxy. This
relationship is actually in good agreement with the
observations, both for single galaxies and for galaxies
in pairs (Zasov 1974; Zasov and Rubtsova 1989;
Zasov and Sulentic 1994; Karachentsev et al. 1999a,
2004).
 To check how universal this conclusion is for
various galaxies and to quantitatively compare $M_{HI}$ and
$M^{c}_{HI}$, below we consider several samples of late-type
galaxies with widely differing parameters - from clumpy irregular
galaxies with bright sites of star formation to Malin-1-like
galaxies of extremely low disk surface brightness with very low
SFRs.

 The optical isophotal diameter is of little use for
this purpose, since it offers no possibility of comparing
galaxies with different surface brightnesses: the
lower the surface brightness of the disk with the same
radial scale length, the smaller its isophotal diameter.
Therefore, it would be more appropriate to compare
galaxies by using the radial disk scale length $R_{0}$
rather than the optical diameter. On average, $R_{HI}\approx 5.4R_{0}$
for both irregular and spiral galaxies (Swaters
et al. 2002). Below, we use this relationship to estimate $M^{c}_{HI}$.

\section{SAMPLES OF GALAXIES}

 In this paper, we consider several different samples
of late-type galaxies with known hydrogen masses $M_{HI}$,
rotational velocities $V_{c}$ (determined in most cases from the
HI line width), and photometric radial scale lengths $R_{0}$:
dwarf irregular (dIrr) galaxies, HI-rich late-type galaxies, UMa
cluster galaxies, clumpy irregular (cIrr) galaxies, which were
included in the atlases of interacting systems by
Vorontsov-Vel'yaminov (1959, 1977) due to their peculiar
appearance, edge-on late-type galaxies, and
low-surface-brightness galaxies including three objects of extreme
Malin-1-type. If required, the distances to galaxies with
significant systemic velocities were reduced to the Hubble
constant $H_{0} = 75$ km s$^{-1}$ Mpc$^{-1}$. For nearby dwarf
galaxies, we used distance taken from original works.

  Let us consider the samples separately.

 (1) Galaxies that morphologically belong to late-type
spirals or irregulars with an absolute magnitude of -18 or
fainter and line-of-sight velocities lower than 3000 km s$^{-1}$
constituted the sample of dwarf galaxies. These are
non-interacting systems: they have no apparent companions within
30$'$ whose velocities differ by less than 500 km s$^{-1}$ from
the velocity of the galaxy. All of the parameters, except the
optical isophotal radius $R_{25}$, were taken from the paper by
van Zee (2001). The radius was calculated from the assumed
distance and the angular diameter $D_{25}$ (the latter was taken
from the LEDA database).

 (2) Late-type galaxies with an absolute magnitude
of -17 or fainter and a high flux density in the HI
line ($S_{HI} >$ 200 mJy) constituted the second sample
of dwarf galaxies. Data for them were taken from the
paper by Swaters et al. (2002).With the exception of
several galaxies, all of them lie at distances less than 25 Mpc.

 (3) Galaxies from the catalog by Vorontsov-
Vel'yaminov (1959, 1977) whose peculiar shapes are very likely to
be attributable not to the system multiplicity, but to the clumpy
distribution of bright star-forming regions constituted the sample
of clumpy irregular (cIrr) galaxies. Several tens of such
galaxies were identified, but the radial disk scale length could
be estimated only for several objects: due to the presence of
bright condensations, the surface brightness of the disks in such
galaxies by no means can be always  described by an exponential
law. Data for these galaxies were taken from the literature
(Patterson and Thuan 1996; Yasuda et al. 1997; Martin 1998;
Bremnes et al. 1999; Iglesias- Paramo and Vilchez 1999; Makarova
1999; Thuan et al. 1999; Barazza et al. 2001; Cairos et al.
2001a, 2001b; Shapley et al. 2001; Pustilnik et al. 2003) and
from the HyperLeda database
(http://www-obs.univ-lyon1.fr/hypercat/).

 (4) Late-type spirals from the UMa open cluster
that differ widely in surface brightness constituted
the sample of normal spiral galaxies. Data for these
galaxies were taken from the paper by Verheijen and
Sancisi (2001). Only the galaxies with the most reliable
measurements that were identified by the authors
as galaxies with fully analyzed data were used.

 (5) The sample of low-surface-brightness galaxies
was taken from the paper by de Block (1996). The central
brightness of the disks in these galaxies is $\mu _{0}(B) \geq
23^{m}/\Box'' $. The sample was supplemented by three objects with
extremely low surface brightnesses: Malin 1, F568-6, and
1226+0105. All of the data for the latters were taken from the
paper by Sprayberry et al. (1993) and reduced to $H_{0} = 75$ km
s$^{-1}$ Mpc$^{-1}$.

 (6) The sample of edge-on spiral galaxies was
drawn from the RFGC catalog (Karachentsev et
al. 1999b). The latest version of this catalog contains
4236 galaxies distributed over the entire sky with
apparent axial ratios $a/b \geq 7$ and angular diameters
$a \geq 0'.6$. Sc-Sd galaxies with small bulges constitute
the bulk of the catalog. The RFGC objects are rich in
gas and are easily detectable in the 21-cm HI line.

No correction for the projection is required for them
when estimating the rotational velocity of the outer
disk. More importantly, the selection criteria make
the sample homogeneous in the structure of its constituent
galaxies. HI data for the galaxies are given in
the paper by Karachentsev and Smirnova (2002). The
radial disk scale lengths for the edge-on galaxies were
taken from the papers by Bizyaev (2000), Bizyaev
and Mitronova (2002), and Kregel et al. (2002).
Since $R_{0}$ determined for the B band was used in
other samples, the disk scale length for the edge-on
galaxies obtained in the near infrared was also
reduced to the B band using the relations

\begin{equation}
R_{0B} = 1.44R_{0I},~~~ R_{0B} = 1.65R_{0K},
\end{equation}

\noindent
 where $R_{0B}$, $R_{0K}$, and $R_{0I}$ are the radial
disk scale lengths in the B, K, and I bands, respectively (de
Grijs 1998).

\section{THE UNIVERSALITY OF THE GAS
MASS-SPECIFIC ANGULAR MOMENTUM
RELATIONSHIP}

 Since the samples are heterogeneous, the $M_{HI}$ and
$R_{0}$ estimates of interest may have various systematic
errors. Nevertheless, all of the samples show similar
relationships, although some of them are noticeably
displaced from one another in the diagrams.

Figure 3 shows the relationships between the neutral hydrogen
mass and the products $R_{25}V_{c}$ (Fig. 3a) and $R_{0}V_{c}$
(Fig. 3b) whose existence follows from the assumption that the
mass of the gas is close (or proportional) to the critical value
for gravitational stability (see the section 2). Both these
products characterize the specific angular momentum of rotating
disks. The diagrams reveal a correlation between the quantities
being compared, which, as would be expected, becomes more
pronounced when using the radial scale length $R_{0}$.

\begin{figure}
\centering
\includegraphics[width=6.5cm, angle=-90 ]{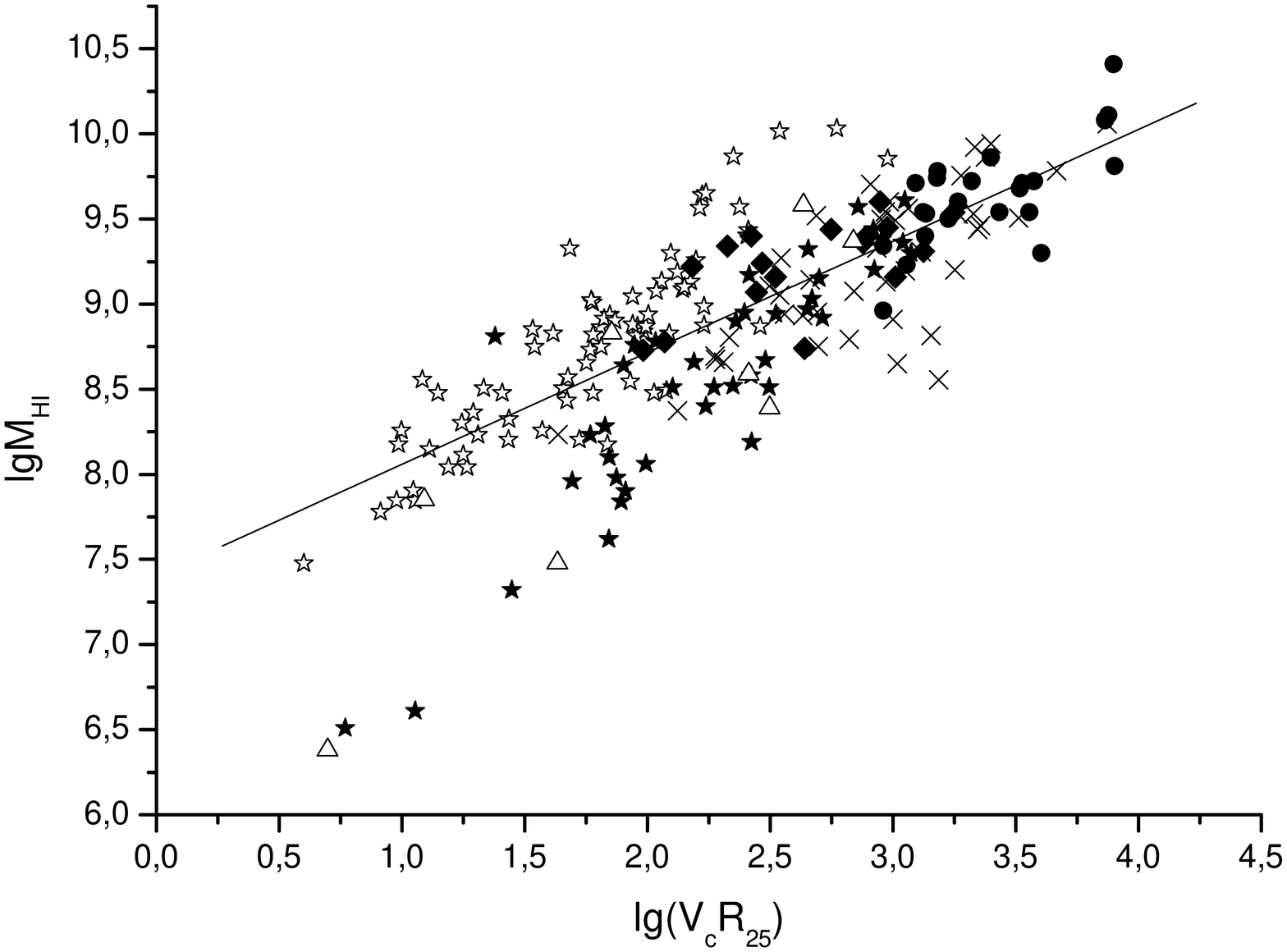}
\includegraphics[width=6.5cm, angle=-90 ]{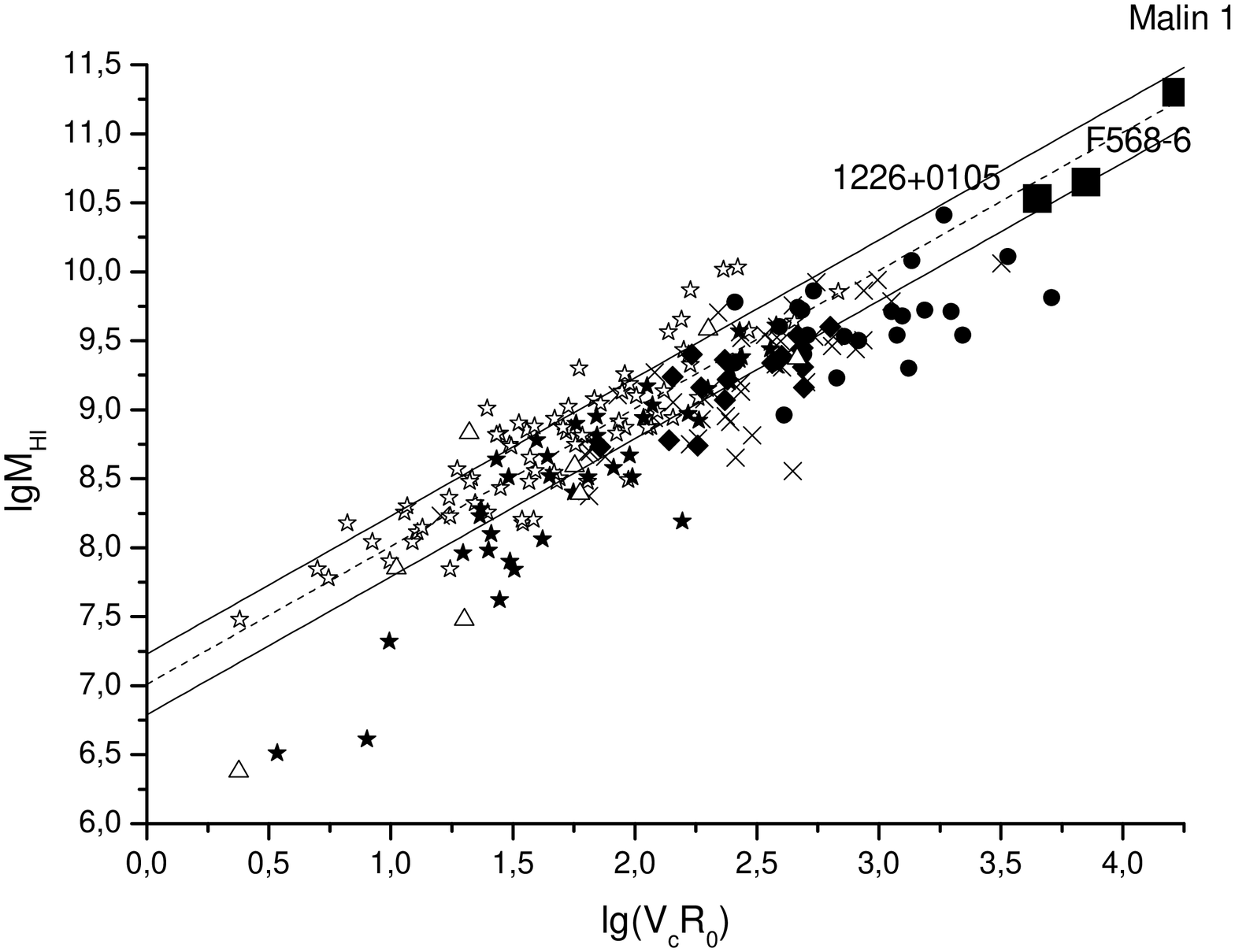}
\caption{HI mass-specific angular momentum diagram for galaxies
of different samples. The logarithm of the product $V_{c}R_{25}$
(top figure) or $V_{c}R_{0}$ (bottom figure) (where $V_{c}$ is the
rotational velocity (km s$^{-1}$), $R_{25}$ is the optical radius
of the galaxy (kpc), and $R_{0}$ is the photometric radial disk
scale length), is along the horizontal axis; the logarithm of the
HI mass (in solar units) is along the vertical axis: (1) dwarf
galaxies with quiescent star formation, (2) hydrogen-rich
galaxies, (3) clumpy irregular (cIrr) galaxies, (4)
low-surface-brightness galaxies, (5) edge-on late-type spirals,
(6) UMa cluster spirals, and (7) three Malin-1 type galaxies.}
\label{array}
\end{figure}

 If these relationships were a simple reflection of the
already discussed size-$HI$ mass relation, then including the
rotational velocity would blur them appreciably. In fact,
although $M_{HI}$ correlates both with $R_{0}$ and (slightly
worse) with $V_{c}$ (Figs. 4a and 4b), the correlation
coefficient between the HI mass and the product of these
quantities is as high (if not higher) as that between $M_{HI}$
and $R_{0}$. It thus follows that the observed relationship
between the HI mass and the angular momentum of the disk does not
reduce to the combination of two simpler relationships, but
reflects the actually existing correlation between the total gas
mass and the kinematic parameters of the gaseous disk.
 The correlation coefficients $r$ and the parameters
of the linear dependences Y = a + b  X considered in this section
are given in the table. The coefficient $b_{c}$ is the mean slope
Y/X of two regression lines: Y(X) and X(Y).

The band bounded by the parallel lines in Fig. 3b describes the
expected relationship between the critical mass $M^{c}_{HI}$ and
$V_{c}R_{0}$ for the case where the total mass of the gas in the
disk (which was assumed here to be proportional to $M_{HI}$) is
close to its critical value: $M_{gas} = M^{c}_{gas}$ (see (6)).
The band width characterizes the uncertainty in the coefficients
in the equation (see the second section) and primarily in the
coefficient $K$, the ratio of the gas velocity dispersion to the
Toomre parameter. The most probable values of $M^{c}_{HI}$
correspond to the dashed line drawn through the center of the
band. As we see from Fig. 3b, the overwhelming majority of points
in the diagram lie within or slightly below the uncertainty band.
The relatively large scatter in the positions of cIrr galaxies
(triangles) and edge-on galaxies (filled circles) in the
log$M_{HI}$-log$(V_{rot}R_{0})$ diagram is probably attributable
to the larger errors in the radial scale lengths of their disks.
The $HI$ mass also deviates significantly from the expected value
for several most slowly rotating gas-poor dwarf galaxies (in the
lower part of the diagram).

Some of the gas-rich dwarf galaxies chosen by a high flux density
in the HI line (open stars) are located slightly above, but also
parallel to the relationship for $M^{c}_{HI}$. The mean gas
surface density in them slightly exceeds the critical value
calculated with the assumed values of $c_{g}/Q_{T}$ (see the
section 2).

 The estimates of the observed hydrogen mass and
the most probable critical mass $M^{c}_{HI}$ (corresponding
to the central line of the band in Fig. 3b) for the galaxies
under consideration are compared in Fig. 5. Like
the diagram in Fig. 3b, the diagram in Fig. 5 clearly
reveals a close correlation between the observed gas
content in the galaxies and the kinematic parameters
of their disks.

\begin{figure}
\centering
\includegraphics[width=6.5cm, angle=-90 ]{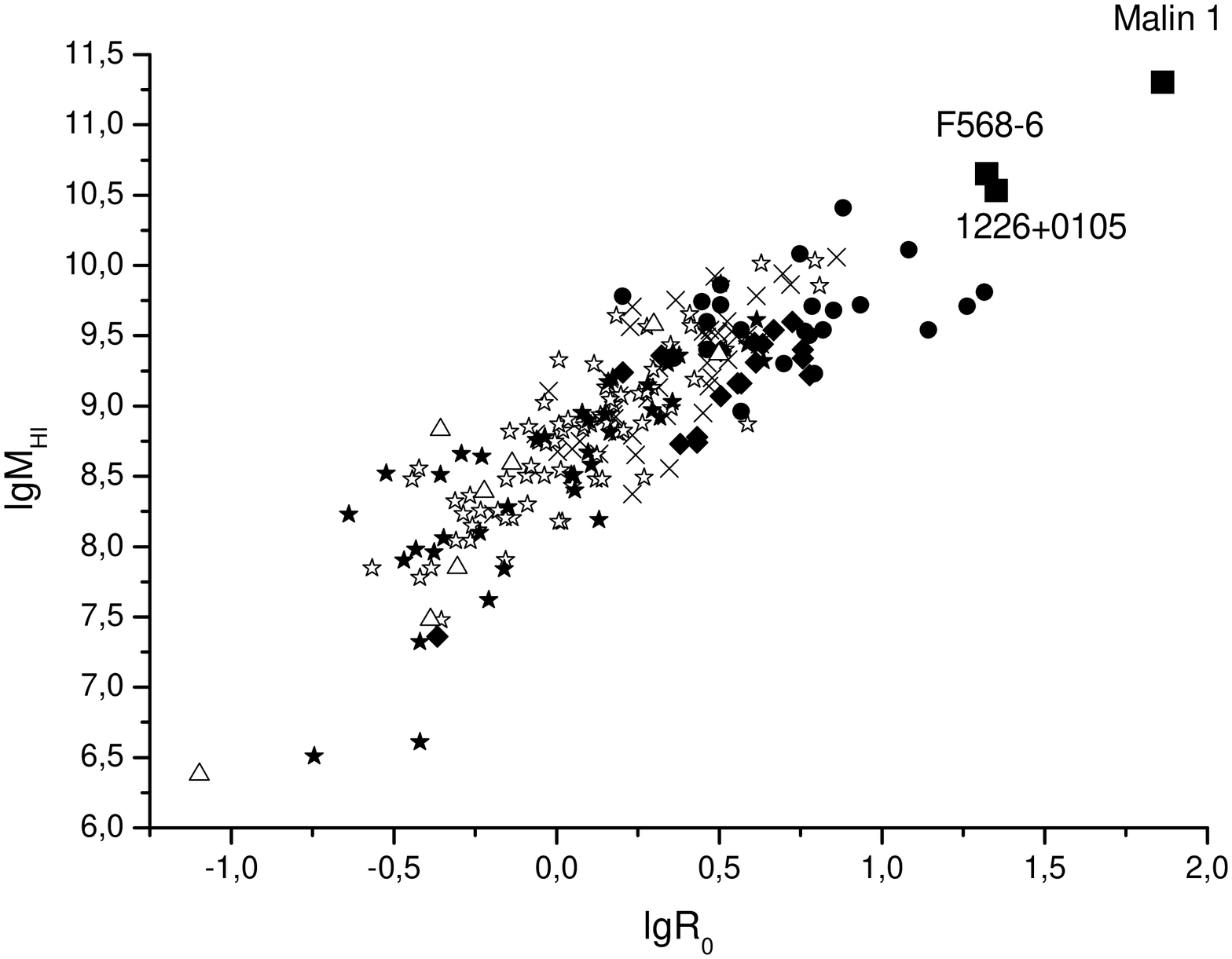}
\includegraphics[width=6.5cm, angle=-90 ]{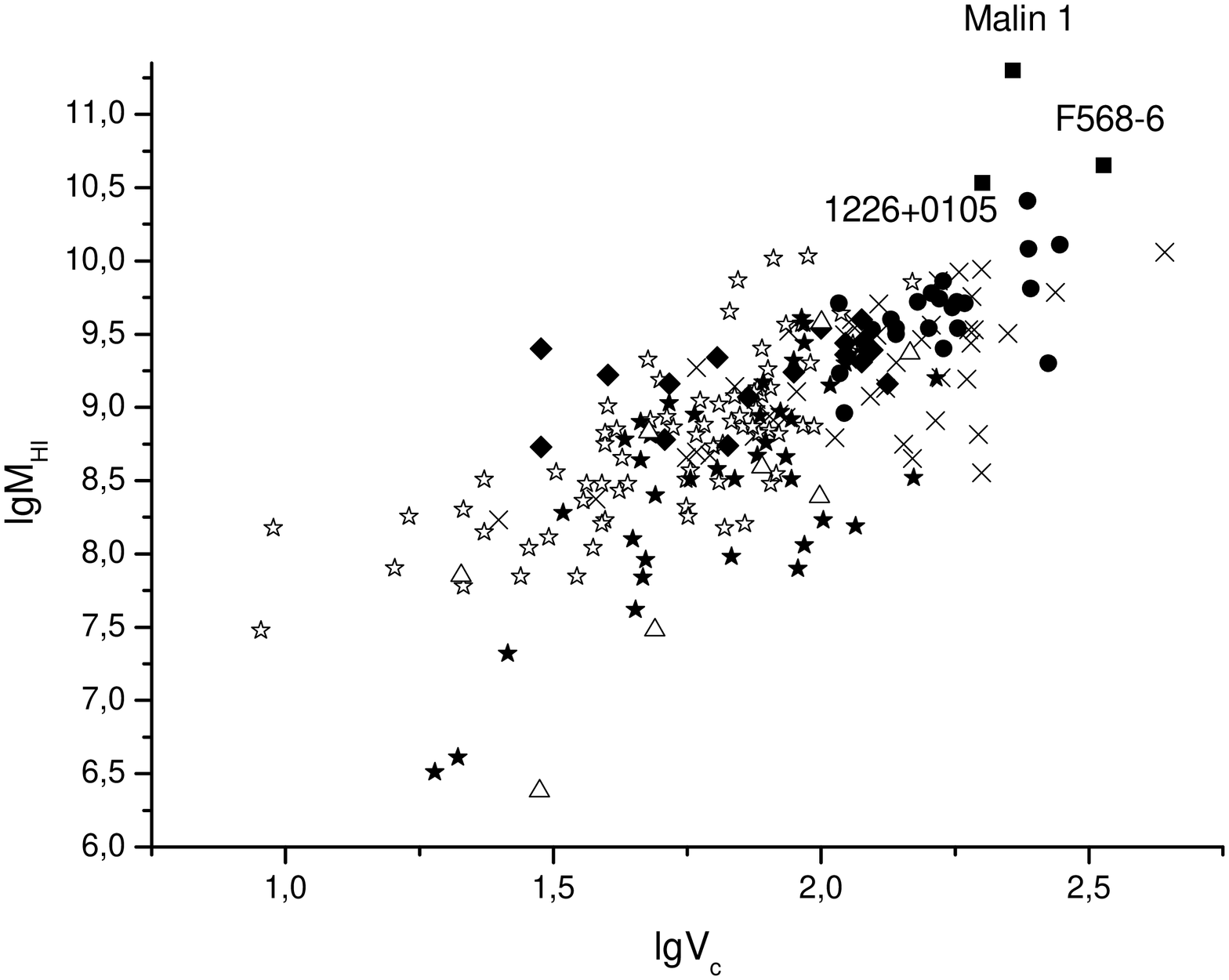}
\caption{(Top) Radial disks scale length $R_{0}$ - HI mass
diagram; (Bottom) rotational velocity $V_{c}$ - HI mass diagram.}
\label{array}
\end{figure}

\section{DISCUSSION AND CONCLUSIONS}
   The agreement between the theoretically expected
(for a marginally stable disk) and observed HI masses (Figs. 3b
and 5), along with the relatively small number of galaxies
located above the highlighted band in Fig. 3b, in which the
hydrogen mass evidently exceeds $M_{c}$, suggest that the
gravitational stability condition for a gaseous layer is an
important factor that determines the amount of gas at the current
epoch and, hence, regulates the star formation efficiency and the
gas depletion rate. If we exclude the galaxies chosen by a high
$HI$ flux density (open asterisk), then, on average, the total HI
mass in the galaxies under consideration proves to be slightly
lower than $M^{c}_{HI}$ (by a factor of 1.5-2). This is probably
the result of a low (compared to the critical value) gas density
in the outer disk regions, beyond $D_{25}$, where a tangible
fraction of the total HI mass is contained (Broeils and Rhee
1997).

\begin{figure}
\centering
\includegraphics[width=6.5cm, angle=-90 ]{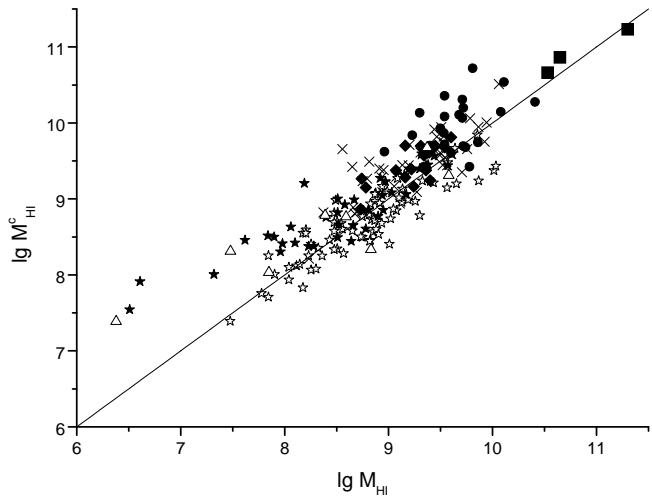}
\caption{(Logarithmic HI mass-critical mass $M^{c}_{HI}$
dependence. The straight line corresponds to the equality between
$M_{HI}$ and $M^{c}_{HI}$.} \label{array}
\end{figure}

 The correspondence of the HI mass to its expected
critical value of  $M^{c}_{HI}$ (and of the total gas mass to its
critical value of $M^{c}_{gas}$) for late-type galaxies can be
naturally explained by assuming that during the evolution of the
galaxy, when the mean gas density decreased to a level close to
the critical one for gravitational stability of the gaseous
layer, the gas depletion slowed down significantly, and, as a
result, most of the galaxies could preserve an amount of gas
close to (or slightly smaller than) $M^{c}_{gas}$ up to the
present time. This conclusion applies to most of the galaxies of
all  the samples considered, including the extremely
low-brightness Malin-1-type galaxies, since they lie on the
general relationship (Fig. 3b). The gas content in these unusual
galaxies proved to be close to the expected values for disks with
enormous angular momentum. The star formation rates and
efficiency in their disks, as well as in the disks of normal
brightness galaxies, must have been decreased when the gas
density reduced below the critical level. However, this does not
rule out the possibility that different initial formation
conditions or a lower initial density of the gaseous disk are
responsible for their low brightness.

The approximate proportionality between the rotational
velocity $V_{c}$ and the size of galaxies that was
pointed out by several authors (Tully and Fisher 1977;
Karachentsev et al. 1999a) can explain why the gas
mass proportional to $V_{c}R_{25}$ changes linearly with
the square of the galactic disk size (the mean HI
surface density is almost constant). The most clear
and almost linear relationship between the rotational
velocity and the optical size of galaxies was found for
a homogeneous sample of nearby galaxies: $logD\sim (0.99 \pm 0.06) log V_{c}$
(Karachentsev et al. 1999a).

 Another important conclusion also follows from
the existence of a linear relationship between the observed mass
$M_{HI}$ and $M^{c}_{HI}$: galaxies with slowly rotating disks
must, on average, possess a higher relative gas mass, i.e., have
a lower ratio of the total (indicative) mass $M_{25}$ within the
photometric radius to the total HI mass. Indeed, since the
photometric radius is proportional to $R_{0}$ and since the total
mass $M_{25} \sim V^{2}_{c} R_{0}$, the relation $M_{25}/M_{HI}
\sim V_{c}$ follows from the condition $M_{HI} \sim V_{c}R_{0}$.
This conclusion is in a good agreement with the observational data
(Fig. 2).

 Thus, the observed gas content in late-type galaxies
reflects a similar (for most of them) pattern of evolutionary
change of the gas mass in the disk. The gravitational instability
of the gaseous layer must play a crucial role in this evolution.
The growth of instability probably facilitated to the enhancement
of star formation and to the fast gas depletion at the initial
(violent) evolutionary stage of the galactic disk. Note that at
the early stage  the density of the gaseous disk exceeded
significantly the critical value calculated for the current gas
velocity dispersion  reflecting the quiescent pattern of star
formation.

\begin{table}
\centering
\caption{Correlation coefficients and parameters of the linear dependences  Y=aX+b} \label{imalog}
\begin{tabular}{@{}lllll@{}}
\hline
        X, Y                  &     r    &        a         &         b        &     b$_{c}$       \\
\hline
log $V_{c}R_{25}$, log$M_{HI}$  &    0.8   & 7.25 $\pm$ 0.10  & 0.71 $\pm$ 0.04   & 0.94 $\pm$ 0.07   \\
log $V_{c}R_{0}$,  log$M_{HI}$  &    0.9   & 7.05 $\pm$ 0.07  & 0.91 $\pm$ 0.03   & 1.05 $\pm$ 0.04   \\
log $R_{0}$, log$M_{HI}$        &    0.9   & 8.63 $\pm$ 0.03  & 1.45 $\pm$ 0.06   & 1.69 $\pm$ 0.08   \\
log $V_{c}$, log$M_{HI}$        &    0.7   & 5.57 $\pm$ 0.22  & 1.79 $\pm$ 0.11   & 2.52 $\pm$ 0.24   \\
 \hline
\end{tabular}
\end{table}

This conclusion should be valid for galaxies with various
diameters, rotational velocities, current SFRs rates, and disk
surface brightnesses.

  In this paper, we have not considered early-type
(S0-Sab) disk galaxies, which contain little gas
at the same sizes and rotational velocities as those
of late-type galaxies. These systems must have a
slightly different star formation history; they have
lost a significant fraction of their gas either through
external factors (e.g., due to the gas being swept up
as it moved in the intergalactic medium of the cluster)
or through internal processes that provided active
star formation even when the gas density decreased
below the critical level for large-scale gravitational
instability.
\bigskip

 \textbf{ACKNOWLEDGMENTS}
\bigskip

We wish to thank I .D. Karachentsev for a discussion of the work
and for presenting the paper (Karachentsev et al. 2004) before
its publication. Authors also thank Dr. G.Galletta for the
electronic version of CISM. This work was supported in part by
the Russian Foundation for Basic Research (project no.
04-02-16518).


\begin{thebibliography}{}
\bibitem{}   F. D. Barazza, B. Binggeli, and P. Prugniel, Astron.
Astrophys. 373, 12 (2001).
\bibitem{}   R. Becker, U. Mebold, K. Reif, et al., Astron. Astrophys.
203, 21 (1988).
\bibitem{}    G. J. Bendo, R. D. Joseph,M.Wells, et al., Astron. J.
124, 1380 (2002).
\bibitem{}   D. Bettoni, G. Galletta, and S. Garcia-Burillo, Astron.
Astrophys. 405, 5 (2003).
\bibitem{}   D. Bizyaev, astro-ph 0007242 (2000).
\bibitem{}    D. Bizyaev and S. Mitronova, Astron.Astrophys. 389,
795 (2002).
\bibitem{}    S. Boissier, A. Boselli, N. Prantzos, et al., MNRAS.
321, 733 (2001).
\bibitem{}   S. Boissier, N. Prantzos, A. Boselli, and G. Gavazzi),
MNRAS. 346, 1215 (2003).
\bibitem{}   A. Boselli, J. Lequeflux, and G. Gavazzi, Astron. Astrophys.
384, 33 (2002).
\bibitem{}   T. Bremnes, B. Binggeli, and P. Prugniel, Astron.
Astrophys. Suppl. Ser. 137, 337 (1999).
\bibitem{}   A. H. Broeils and M.-H. Rhee, Astron. Astrophys.
324, 877 (1997).
\bibitem{}   L. Cairos, N. Caon, J. Vilchez, et al., Astrophys.
J. Suppl. Ser. 136, 393 (2001b).
\bibitem{}  L. Cairos, J. Vilchez, J. Gonzalez Perez, et al., Astrophys.
J. Suppl. Ser. 133, 321 (2001a).
\bibitem{}   F. Casoli, S. Sauty, M. Gerin, et al.), Astron. Astrophys.
331, 451 (1998).
\bibitem{}   F. Combes and J.-F. Becquaert, Astron. Astrophys.
326, 554 (1997).
\bibitem{}   F. Combes and D. Pfenniger, Astron. Astrophys. 327,
453 (1997).
\bibitem{}   W. J.G. de Block ,S. S. McGaugh, and J.M. van der
Hulst, MNRAS 283, 18 (1996).
\bibitem{}   R. de Grijs, MNRAS 299, 595 (1998).
\bibitem{}   N. Devereflux and S. Hameed, Astron. J. 113, 599
(1997).
\bibitem{}   B. G. Elmegreen,MNRAS 275, 944 (1995).
\bibitem{}   J. Hewitt, M. Haynes, and R. Giovanelli, Astron. J.
88, 272 (1983).
\bibitem{}   D. A. Hunter, B. G. Elmegreen, and A. L. Baker,
Astrophys. J. 493, 595 (1998).
\bibitem{}   J. Iglesias-Paramo and J. M. Vilchez, Astrophys. J.
518, 94 (1999).
\bibitem{}   I. D. Karachentsev, V. E. Karachentseva,
W. K. Huchtmeier, et al., Astron. J. 127, 2031
(2004).
\bibitem{}   I. Karachentsev, V. Karachentseva, Y. Kudrya, et al.,
Bull. SAO 47, 5 (1999b).
\bibitem{}   I. D. Karachentsev, D. I. Makarov, and W. K. Huchtmeier,
Astron. Astrophys. Suppl. Ser. 139, 97
(1999a).
\bibitem{}   I. D. Karachentsev and A. V. Smirnova, Astrofiz. 45,
448 (2002).
\bibitem{}   W.-T. Kim and E. Ostrik er, Astrophys. J. 559, 70
(2001).
\bibitem{}   T. Kranz, A. Slyz, and H. W. Rix, Astrophys. J. 586,
143 (2003).
\bibitem{}   M. Kregel, P. van der Kruit, and R. de Grijs, MNRAS
334, 646 (2002).
\bibitem{}   R. Larson and B. Tinsley, Astrophys. J. 219, 46
(1978).
\bibitem{} B. Lewis, Astrophys. J. 285, 453 (1984).
\bibitem{} L. Makarova, Astron. Astrophys.Suppl. Set. 139, 491
(1999).
\bibitem{} M.C. Martin, Astron. Astrophys. Suppl. Ser. 131, 77
(1998).
\bibitem{} C. Martin and R. Kennicutt, Astrophys. J. 555, 301
(2001).
\bibitem{} S. McGaugh S. and W. J. G. de Blok, Astrophys. J.
481, 689 (1997).
\bibitem{} A. G. Morozov, Sov. Astron. 29, 120 (1985).
\bibitem{} R. Patterson and T. Thuan, Astrophys. J. Suppl. Ser.
107, 103 (1996).
\bibitem{} D. L. Polyachenko, E. V. Polyachenko, and
A. V. Strel'nikov, Astron. Lett. 23, 483 (1997).
\bibitem{} S. Pustilnik, A. Zasov, A. Kniazev, et al.), Astron.
Astrophys. 400, 841 (2003).
\bibitem{} W. Quirk, Astrophys. J. Lett. 176, L9 (1972).
\bibitem{} A. Shapley, G. Fabbiano, and P. Eskridge, Astrophys.
J. Suppl. Ser. 137, 139 (2001).
\bibitem{} E. J. Shaya and S. R. Federman, Astrophys. J. 319,
76 (1987).
\bibitem{} D. Sprayberry, C. D. Impey,M. J. Irwin, et al., Astrophys.
J. 417, 114 (1993).
\bibitem{} R. A. Swaters, T. S. van Albada, J. M.van der Hulst,
and R. Sancisi, Astron. Astrophys. 390, 829 (2002).
\bibitem{} T. X. Thuan, V. A. Lipovetsky, J.-M. Martin, and
S. A. Pustilnik), Astron. Astrophys. Suppl. Ser. 139,
1 (1999).
\bibitem{} R. B. Tully and J. R. Fisher, Astron. Astrophys. 54,
661 (1977).
\bibitem{} M. A. W. Verheijen and R. Sancisi, Astron. Astrophys.
370, 765 (2001).
\bibitem{} L. van Zee, Astron. J. 121, 2003 (2001).
\bibitem{} B. A. Vorontsov-Velyaminov, Atlas and Catalogs of Interacting Galaxies (Mosk. Gos. Univ, Moscow,
1959)[in Russian].
\bibitem{} B. A. Vorontsov-Velyaminov, Astron. Astrophys. 28, 1 (1977).
\bibitem{} T. Wong and L. Blitz, Astrophys. J. 569, 157 (2002).
\bibitem{} N. Yasuda,M. Fukugita, and S.Okamura, Astrophys. J. Suppl. Ser. 108, 417 (1997).
\bibitem{} A. V. Zasov, Sov. Astron.18, 730 (1974).
\bibitem{} A. V. Zasov, Astron. Lett. 21, 652 (1995).
\bibitem{} A. V. Zasov and D. V. Bizyaev, Publ. Astron. Soc.Pacic 66, 73 (1994).
\bibitem{} A. V. Zasov, A. V. Khoperskov, and N. V. Tyurina, Astron. Lett. 30, 593 (2004).
\bibitem{} A. V. Zasov and T. V. Rubtsova, Sov. Astron. Lett. 15, 51 (1989).
\bibitem{} A. V. Zasov and S. G. Simakov, Astroz. 29, 518(1989).
\bibitem{} A. V. Zasov and J. Sulentic, Astrophys. J. 430, 179
(1994).

\end{thebibliography}
\end{document}